# De-scattering with Excitation Patterning (DEEP) Enables Rapid Wide-field Imaging Through Scattering Media

Dushan N. Wadduwage[1,2,3]*, Jong Kang Park[2,3]*, Josiah R. Boivin[4], Yi Xue[3,5] and Peter T.C. So[2,3,5]

*These authors contributed equally to this work.
[1]Center for Advanced Imaging, Faculty of Arts and Sciences, Harvard University, Cambridge, MA 02138, USA
[2]Dept. of Biological Engineering, [3]Laser Biomedical Research Center, [4]Picower Institute for Learning and Memory,
[5]Dept. of Mechanical Engineering,
Massachusetts Institute of Technology, 77 Massachusetts Ave., Cambridge, MA 02139, USA

**From multi-photon imaging penetrating millimeters deep through scattering biological tissue, to super-resolution imaging conquering the diffraction limit, optical imaging techniques have greatly advanced in recent years. Notwithstanding, a key unmet challenge in all these imaging techniques is to perform rapid wide-field imaging through a turbid medium. Strategies such as active wave-front correction and multi-photon excitation, both used for deep tissue imaging; or wide-field total-internal-refection illumination, used for super-resolution imaging; can generate arbitrary excitation patterns over a large field-of-view through or under turbid media. In these cases, throughput advantage gained by wide-field excitation is lost due to the use of point detection. To address this challenge, here we introduce a novel technique called <u>De</u>-scattering with <u>E</u>xcitation <u>P</u>atterning, or 'DEEP', which uses patterned excitation followed by wide-field detection with computational imaging. We use two-photon temporal focusing (TFM) to demonstrate our approach at multiple scattering lengths deep in tissue. Our results suggest that millions of point-scanning measurements could be substituted with tens to hundreds of DEEP measurements with no compromise in image quality.**

Point-scanning two-photon microscopy (PSTPM) is used routinely for *in vivo*, volumetric biological imaging, especially in deep tissues[1-3]. Imaging of cortical vasculature in mouse brain has been demonstrated down to 1.6 mm[3]. The near-infrared (NIR) or short-wave infrared (SWIR) femtosecond laser pulses, used in PSTPM, have long penetration depth in tissue due to the strong inverse relationship between light scattering and wavelength. The excitation light of PSTPM is focused at the diffraction limit, enabling efficient two-photon excitation. Emission photons from the focal spot, scattered or not, are then collected by a point detector, such as a photomultiplier tube, and assigned to a single pixel of the image. Despite the excellent penetration depth, a conventional PSTPM is slow, due to the need for raster scanning, and imaging time scales linearly as volume increases (see simulation results in Figure 1A), limiting studies of fast biological dynamics.

An attractive alternative to PSTPM is wide-field two-photon microscopy, typically called temporal focusing microscopy (TFM). As the name suggests, TFM achieves optical sectioning by focusing a beam temporally while maintaining wide-field illumination[4-6]. In TFM, wide-field excitation is enabled by femtosecond laser pulses with high pulse energy (~µJ – mJ). Depth resolution is achieved by controlling optical dispersion so that the pulse width rapidly broadens away from the focal plane, resulting in low two-photon excitation efficiency[4]. The optical dispersion profile is controlled by placing a grating on a conjugate image plane at the excitation path. TFM provides excellent penetration of multiphoton excitation light through scattering media. Penetration depths up to seven scattering lengths have been demonstrated with two-photon excitation[7]. With three-photon excitation, we have previously demonstrated up to 700µm penetration through scattering brain tissue[8]. We have also demonstrated optogenetic stimulation using three-photon temporal focusing excitation[8]. In TFM however, emitted photons, with their relatively short wavelengths (compared to excitation photons), are strongly scattered by the tissue. To maintain wide-field imaging, the emission light from the focal plane is typically recorded by a camera. As a result of wide-field detection, some scattered emission photons may be assigned to incorrect pixels on the detector, resulting in degradation of both resolution and signal-to-noise ratio. Consequently, at shallower image planes, TFM images show a background haze (see simulation results in Figure 1B); as the imaging depth is increased, TFM images lose their high-resolution information (see supporting materials).

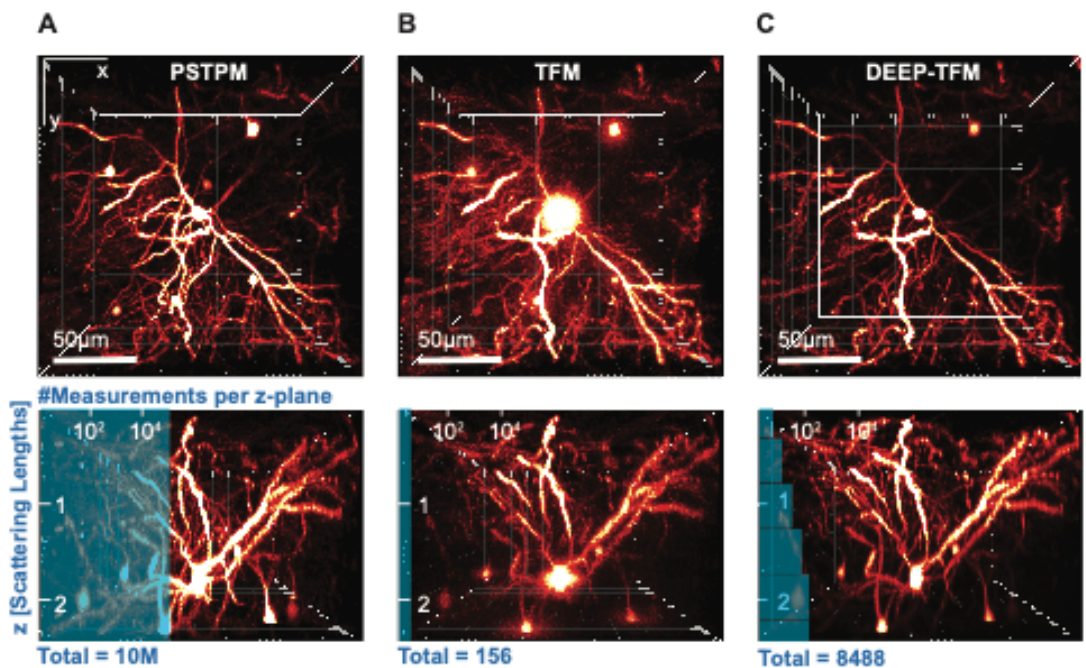

**Figure 1.** Comparison of two-photon imaging scenarios of a whole neuron (256 × 256 × 156 voxels) in a live mouse using: **(A)** PSTPM (experimental), **(B)** TFM (simulated), and **(C)** DEEP-TFM (simulated). Shown on the top row are the top views (XY-view) of the image stacks; shown in the background of the bottom row are the side views (XZ-view). The blue-shaded plots on the bottom rows show the number of measurements needed at each z-plane. PSTPM requires over 10 million measurements (one for each voxel). A conventional TFM requires only 156 measurements (one for each depth), but the image quality degrades as the imaging depth is increased. DEEP-TFM requires 8488 measurements, but maintains similar image quality as PSTPM.

To de-scatter wide-field TFM images, computational imaging approaches have previously been proposed[7,9-11]. While they improve image resolution and contrast at shallower depths (<1 scattering length) most have not been applied for deep imaging. Including our team, a number of groups are working on wide-field TFM for deep imaging[7,11,12]. Escobet-Montalbán and co-workers[7] demonstrated a method called 'TempoRAl Focusing microscopy with single-pIXel detection' or TRAFIX. They used a set of two-dimensional illumination patterns along with single pixel detection[13] to image as deep as seven scattering lengths through a scattering phantom[7]. However, they require the same number of illumination patterns (and measurements) as the number of pixels in the imaged field-of-view; thus no evident speed-up over PSTPM was demonstrated. Alemohammad et. al. also used a similar technique, but with compressive imaging (called compressive temporal focusing two photon or CS-TFTP)[11], demonstrating 10x speed-up over PSTPM. Both TRAFIX and CS-TFTP, however, use point (or single pixel) detection; so their acquisition times are field-of-view (FOV) dependent. Theoretical speed-up over PSTPM, therefore, is strictly limited to compressibility of the image. To the best of our knowledge, no computational TFM with wide-field detection has been demonstrated.

In this paper, we introduce a novel computational de-scattering technique called Descattering by Excitation Patterns in TFM or 'DEEP-TFM'. Similar to previous approaches, we use wide-field temporal focusing patterned excitation; however, the signal is measured with a wide-field detector, such as a camera. Briefly, we built a modified temporal focusing microscope that projects arbitrary excitation patterns onto the focal plane using a digital mirror device (DMD). Emission light from the modulated excitation is then detected by a camera (see Figure 2A and the methods section for a detailed description of the microscope). Due to their NIR wavelengths, the excitation patterns maintain their fidelity despite travelling through a scattering medium[8]. However, the emission photons are scattered by the tissue, and the strength of scattering is strongly depth-dependent. This holds for most biological tissues[8,14]. In practice, TFM images are minimally affected by scattering at or near the surface; as the imaging depth increases, scattering gradually degrades only the high-frequency information in the images. Low frequencies in the images are retained for most depths even with wide-field detection. Single pixel detection approaches discard this low frequency information, and hence require a large number of excitation points. DEEP-TFM combines the information about the excitation patterns (through a calibration process explained in the methods section) with the acquired images, to computationally reconstruct a de-scattered image (see Figure 2B). Experimentally, to de-scatter a single FOV, multiple patterned excitations (and images) are needed; the number depends on the loss of high-frequency information due to scattering, and hence on the imaging depth.

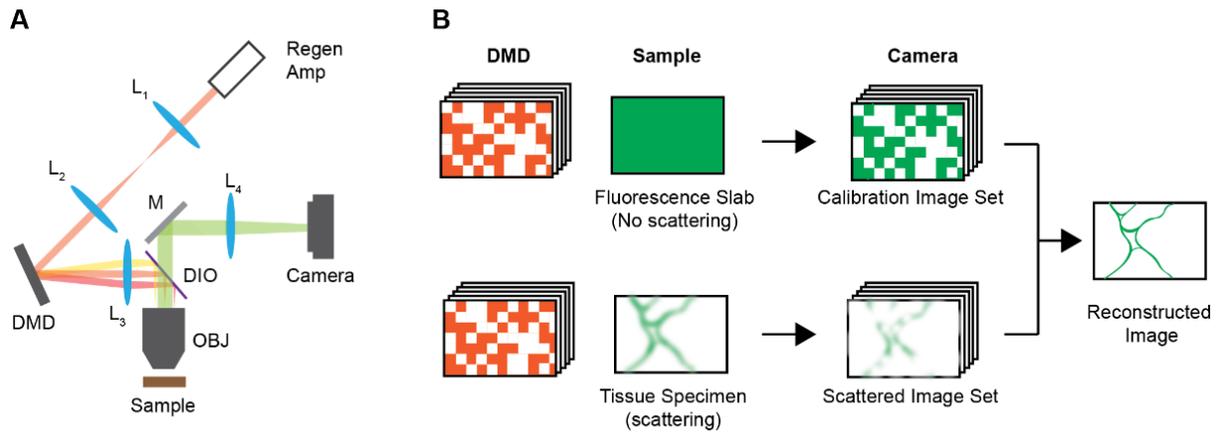

**Figure 2. (A)** Optical schematic of the imaging system: $L_1$, $L_2$, $L_3$, $L_4$ – lenses; DMD- Digital Mirror Device; DIO – Dichroic mirror; OBJ – Microscope objective. **(B)** Proposed computational imaging strategy. First, a set of patterns are projected on a calibration specimen (homogeneous thin fluorescent layer) to record the calibration image set in the absence of any scattering. Then, the same patterns are projected to record the encoded images through a scattering medium. The de-scattered images are then reconstructed.

Mathematically, the DEEP-TFM imaging process can be modelled by the following equation.

$$Y_t(x,y) = sPSF(x,y) * \left\{ \left( exPSF(x,y) * \widetilde{H}_t(x,y) \right) \circ X(x,y) \right\} \quad \text{(Eq1)}$$

Here $x$ and $y$ are spatial coordinates; $t$ is time; $exPSF(x,y)$, and $sPSF(x,y)$ are the excitation and scattering point spread functions; $\widetilde{H}_t(x,y)$ is the $t^{th}$ modulation pattern projected by the DMD. $X(x,y)$ is the object being imaged; $Y_t(x,y)$ is the $t^{th}$ image acquired on the camera. The operators $*$ and $\circ$ represent spatial convolution and pixel wise multiplication (see supporting section S1 for more details). Writing Eq1 in the spatial Fourier domain we get,

$$\mathcal{F}Y_t(kx,ky) = \mathcal{F}sPSF(kx,ky) \circ \left\{ \left( \mathcal{F}exPSF(kx,ky) \circ \mathcal{F}\widetilde{H}_t(kx,ky) \right) * \mathcal{F}X(kx,ky) \right\} \quad \text{(Eq2)}$$

Assuming $N$ pixels (both in the image, $Y_t$, and the object, $X$), the above equation has $2N$ unknowns ($N$ in $\mathcal{F}X$, and $N$ in $\mathcal{F}sPSF$). As written, each measurement (i.e. an image taken at time point $t$) appears to provides $N$ equations. However, $\mathcal{F}sPSF$ acts as a low-pass filter and for out-band frequencies (out of the frequency support of $\mathcal{F}sPSF$) the right-hand side of the Eq2 is zero. Now, assume that $\mathcal{F}sPSF$'s frequency support has $M$ pixels. Then, each measurement provides $M$ independent equations. Thus, for the above system of equations to be solvable, we need $N_t > 2N/M$ measurements. For deep tissue imaging applications, the frequency support of $\mathcal{F}sPSF$ changes with imaging depth. As there is negligible scattering at the surface, at the surface $M \sim N$; we only need $N_t = O(1)$ measurements (here '$O(.)$' represents the asymptotic 'big-O' notation). At very deep frames where there is no spatial information on the recorded images, $M \sim 1$. We hence need $N_t = O(N)$ measurements. Please also note that because of the frequency domain convolution between $\mathcal{F}\widetilde{H}_t$ and $\mathcal{F}X$ (Eq2) out-band frequencies (of the frequency support of $\mathcal{F}sPSF$) in $X$ are still sampled on to $Y_t$ as long as $\mathcal{F}\widetilde{H}$ captures all possible frequencies of $X$. It can be shown that a random pattern of $\widetilde{H}$ satisfies this criterion (see supporting Figure S1 and supporting section S1). Thus, an ensemble of $O(2N/M)$ random patterns, $\{\widetilde{H}_t\}$, can be used to fully measure $X$ in DEEP-TFM (see supporting section S1 for a detailed description). Upon such measurement we record an ensemble of $\{Y_t\}$ images corresponding to $\{\widetilde{H}_t\}$; $X$ can be reconstructed using, $\{Y_t\}$ and $\{H_t\}$, by solving Eq1 or its corresponding frequency domain representation, i.e. Eq2 (see the methods section and supporting section S1 for a detailed description).

Figure 3 shows representative examples of DEEP-TFM imaging. We first imaged a mixture of $4\ \mu m$ and $10\ \mu m$ beads through $2\ mm$ of a scattering lipid solution (0.15%). Figure 3A1 shows a conventional TFM image; Figure 3A2 shows the final DEEP-TFM image reconstructed with $N_t = 128$ measurements. Since the light from all the beads goes through the same thickness of scattering medium, one would expect all $4\ \mu m$ beads to show similar scattering behavior in Figure 3A1, as the thickness of the TFM excitation plane is around $15\ \mu m$ (supporting Figure SX). In fact they do; some beads only appear to show more scattering because they are out of the focal plane. In the DEEP-TFM image, the out of focus beads are not visible. Thus, in addition to de-scattering, DEEP-TFM also improves on the axial resolution of TFM, to what extent still needs to be determined.

Next, we imaged a $16\mu m$ thick mouse kidney section through the same $2\ mm$ of scattering lipid solution (Figure B1-2). The inset of Figure 3B1 shows a representative patterned excitation. illustrating the immediate improvement in image contrast and signal to background ratio achieved by DEEP-TFM (see supporting Figure S2 for additional results). We also imaged a $200\mu m$ thick section of muscle tissue stained for nuclei (blue channel, Hoechst 33342) and F-actin (red channel, Alexa Fluor 568 Phalloidin). The FOV was nearly $150 \times 150\ \mu m^2$ with $256 \times 256$ pixels. All DEEP-TFM reconstructions were performed with $N_t = 128$ measurements. Representative images comparing TFM and DEEP-TFM at a 190 $\mu m$ deep plane are shown in Figures 3C1-2. Finally, in Figure 3D, we show a direct comparison of TFM and DEEP-TFM in the same F-actin image (at a 170 $\mu m$ deep plane). As seen, in deep imaging, TFM loses a significant amount of high-frequency information, with almost no high-resolution detail visible. In contrast, with DEEP-TFM, most fine details are reconstructed.

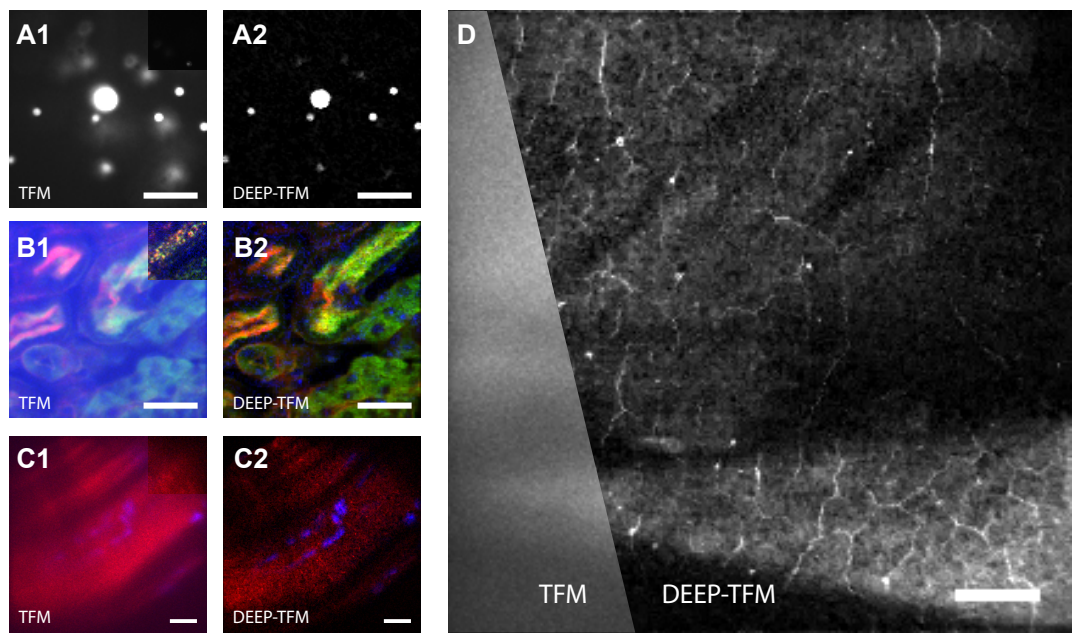

**Figure 3. (A1)** Wide-field temporal focusing two photon microscopy (TFM) image of a mixture of $4\mu m$ and $10\mu m$ beads imaged through a scattering medium. **(A2)** DEEP-TFM image of the FOV in 'A1' reconstructed with 128 measurements ($Nt = 128$). **(B1-2)** Respectively the TFM, and DEEP-TFM (with $Nt = 128$) images of a mouse kidney specimen. Shown in blue, green, and red channels are respectively: nucleus, Alexa Fluor 488 conjugated WGA, and F-actin. **(C1-2)** Respectively the TFM, and DEEP-TFM (with $Nt = 128$) images of a mouse muscle specimen at a $190\mu m$ deep imaging plane. The blue and red channels are respectively: nucleus (stained with Hoechst 33342) and F-actin (stained with Alexa Fluor 568 Phalloidin). **(D)** A representative image of F-actin (the red-channel) of the same sample in 'C' at a $170\mu m$ deep imaging plane comparing the TFM vs. DEEP-TFM (with $Nt = 128$) images. (Scale bars in A1-B2 and C1-D are $30\mu m$ and $20\mu m$, respectively, in length; insets in A1, B1, and C1 show a representative region modulated with a representative pattern).

Our simulation (Figure 1 and supporting section S2) and experimental (Figure 3) results demonstrate the potential of DEEP-TFM as a wide-field multiphoton microscopy approach. DEEP-TFM can resolve deep tissue biological images with similar quality to PSTPM at high resolution. Lateral resolution of DEEP-TFM is determined by the maximum resolution of the projected patterns and hence matches PSTPM. The axial resolution of TFM is a safe upper bound for the axial resolution of DEEP-TFM. However, structured illumination has previously been used to improve the axial resolution of TFM[9,10]. Our results also suggest that we see an improvement in axial resolution (Figure 3A1-2). To the best of our knowledge, DEEP-TFM is currently the only computational wide-field multiphoton imaging method whose frame rate is independent of the FOV size; theoretically, millimeter large FOVs at diffraction limited resolution may be achieved with no sacrifice of speed. Further, DEEP-TFM uniquely provides flexible, depth-dependent imaging speeds: shallow imaging is almost single-shot, while deep imaging speed is depth-optimized. Thus, in theory an increase in speed of more than three orders of magnitude over PSTPM may be achieved for a volume of $256 \times 256 \times 156\ px^3$, with the assumption of the same acquisition time per measurement (see Figure 1). Lastly, DEEP-TFM satisfies all the requirements in modern compressive sensing theory[13]. With suitable image priors, an additional $\sim 10 \times$ increase in speed may be achieved[11].

There are practical limitations to DEEP-TFM. Custom made objective lenses may be needed for a large FOV at a high numerical aperture; total average power causing heat damage in tissue may also limit FOV size (demonstrations presented here use only $\sim 15\ mW$ over a $\sim 150 \times 150\ \mu m^2$ FOV - by comparison TRAFIX used more than 500mW over a $\sim 90 \times 90\ \mu m^2$ FOV[10]); wide-field detection may limit the maximum frame rate readable with current electronics.

In summary, here we present DEEP-TFM (Descattering by Excitation Patterns in Temporally Focused Microscopy), a novel computational wide-field technology for deep tissue multiphoton microscopy. Our results suggest that DEEP-TFM can resolve images with similar quality to point scanning two-photon microscopy, but in a wide-field design; the acquisition time, therefore, is FOV independent. We believe, with optimized instrumentation and compressive data acquisition, DEEP-TFM can accelerate volumetric multiphoton imaging by three to four orders of magnitude.

**Acknowledgments:** We thank Kendyll Burnell for technical assistance with tissue sample preparation. This work was supported by F32 MH115441 (JB), and Center for Advanced Imaging at Harvard University (DNW).

# Reference

[1] W. Denk, J. H. Strickler, and W. W. Webb, "Two-photon laser scanning fluorescence microscopy," Science 248, 73–76 (1990).
[2] F. Helmchen and W. Denk, "Deep tissue two-photon microscopy," Nat. Methods 2, 932–940 (2005).
[3] D. Kobat, N. G. Horton, and C. Xu, "In vivo two-photon microscopy to 1.6-mm depth in mouse cortex," J. Biomed. Opt. 16, 106014 (2011).
[4] D. Oron, E. Tal, and Y. Silberberg, "Scanningless depth-resolved microscopy," Opt. Express 13, 1468–1476 (2005).
[5] L.-C. Cheng, C.-Y. Chang, C.-Y. Lin, K.-C. Cho, W.-C. Yen, N.-S. Chang, C. Xu, C. Y. Dong, and S.-J. Chen, "Spatiotemporal focusing-based widefield multiphoton microscopy for fast optical sectioning," Opt. Express 20, 8939–8948 (2012).
[6] A. Vaziri and C. V. Shank, "Ultrafast widefield optical sectioning microscopy by multifocal temporal focusing," Opt. Express 18, 19645–19655 (2010).
[7] A. Escobet-Montalbán, R. Spesyvtsev, M. Chen, W. A. Saber, M. Andrews, C. Simon Herrington, M. Mazilu, and K. Dholakia, "Wide-field multiphoton imaging through scattering media without correction," Sci Adv 4, eaau1338 (2018).
[8] Rowlands, Christopher J., et al. "Wide-field three-photon excitation in biological samples." *Light: Science & Applications* 6.5 (2017): e16255.
[9] H. Choi, E. Y. S. Yew, B. Hallacoglu, S. Fantini, C. J. R. Sheppard, and P. T. C. So, "Improvement of axial resolution and contrast in temporally focused widefield two-photon microscopy with structured light illumination," Biomed. Opt. Express 4, 995–1005 (2013).
[10] Y. D. Sie, C.-Y. Chang, C.-Y. Lin, N.-S. Chang, P. J. Campagnola, and S.-J. Chen, "Fast and improved bioimaging via temporal focusing multiphoton excitation microscopy with binary digital-micromirror-device holography," J. Biomed. Opt. 23, 1–8 (2018).
[11] Alemohammad, Milad, et al. "Widefield compressive multiphoton microscopy." *Optics letters* 43.12 (2018): 2989-2992.
[12] D. N. Wadduwage, J. K. Park, and P. T. C. So, Proc. SPIE 10499, 1049933 (2018).
[13] Duarte, Marco F., et al. "Single-pixel imaging via compressive sampling." *IEEE signal processing magazine* 25.2 (2008): 83-91.
[14] Rowlands, Christopher J., et al. "Objective, comparative assessment of the penetration depth of temporal-focusing microscopy for imaging various organs." *Journal of biomedical optics* 20.6 (2015): 061107
[15] Bioucas-Dias, José M., and Mário AT Figueiredo. "A new TwIST: two-step iterative shrinkage/thresholding algorithms for image restoration." *IEEE Transactions on Image processing* 16.12 (2007): 2992-3004.

## Methods

**Pattern-illuminated temporally focused wide-field two-photon microscope**
Figure 2A shows the schematic diagram of a temporal focusing microscope that enables arbitrary patterned illumination. First, an ultrafast pulsed laser beam [800 nm center wavelength, 120 fs pulse width, 10 kHz repetition rate, ~8 mm beam diameter ($1/e^2$)] from a regenerative amplifier (Legend Elite, Coherent, Santa Clara, CA, USA) was magnified to ~32 mm and directed to a digital micromirror device, DMD, (DLP LightCrafter 9000 EVM, Texas Instruments, TX, USA). The DMD was used as a diffractive element and pattern generator, simultaneously. The beam was diffracted from the DMD with an effective grating period of ~190 lines/mm with an incident angle of 26.4⁰. Arbitrary patterns could be uploaded onto the DMD using a control program (DLP LCR 9000 GUI) provided by Texas Instruments. The DMD was followed by a 4f-lens system consisting of two planoconvex lenses ($L_1$; $f$ = 250 mm; LA1461, Thorlabs and $L_2$; $f$ = 125 mm; AC254-125-B-ML, Thorlabs, Newton, NJ, USA), which served to project and magnify the image of the DMD. Then the images formed by $L_1$ and $L_2$ were relayed onto the sample, S, through another tube lens ($L_3$; $f$ = 300 mm; AC508-300-B-ML, Thorlabs), and an objective lens, OBJ, (water immersion 20×/1.0, Zeiss, Jena, Germany). The system magnification is about 73× according to the focal lengths of tube lenses and the effective focal length of the objective lens. The geometric dispersion of the system ensured that the pulse width was broad enough to minimize multiphoton excitation except at the sample plane. The objective location was controlled with an objective piezo positioner (MIPOS-500, Piezosystem Jena, Jena, Germany).

The two-photon excited fluorescence from the sample, S, was collected by the same objective lens, OBJ, in an epi detection geometry and reflected by a dichroic filter, DIO (Di03-R635-t3, Semrock, Rochester, NY, USA) to a camera. Then fluorescence signals were imaged by another tube lens, $L_4$, ($f$ = 200 mm; PAC064, Newport, Irvine, CA, USA) onto an EMCCD camera, (iXon+, Andor, Belfast, Northern Ireland). For multicolor detection, three combinations of filter sets were used; blue channel centered at 460 nm (Semrock FF01-460/60-25 and Chroma E530SP-SPC), green channel centered at 535 nm (Chroma ET535/70M and Chroma ET680SP-2P8), and red channel centered at 605 nm (Chroma ET605/70M and Chroma E700SP-2P). An achromatic doublet lens pair (1:2, MAP1050100-A, Thorlabs) was used to expand the image size onto the camera when DMD pixels of 1024 × 1024 were used for pattern generation. For patterns of larger pixel size (1600 × 1600), a 1:1 achromatic doublet lens pair (MAP107575-A, Thorlabs) was used to ensure that the image fit the detector size.

Data from the camera was transferred using either a control program (Andor Solis) provided by Andor or a control software implemented using LabVIEW 2015 (National Instruments, Austin, TX, USA).

**Preparation of the calibration and the scattering Samples**
A thin quantum dot layer was used for the calibration of patterns for the green (535 nm) and red channels (605 nm). A thin, fluorescent layer of green quantum dots (supplied by QDVision, Lexington, MA, USA) dispersed in hexane (10 μL) were dropped onto a coverslip (thickness 170 μm) and allowed to dry. The coverslip was affixed to a glass slide and sealed by transparent nail varnish. A thin DAPI solution layer was used for the calibration of patterns for the blue channel (460 nm). Saturated DAPI solution in 1:1 mixture of deionized water and DMSO were dropped in a pre-holed spacer (120 μm thick, Secure-Seal Imaging Spacers, Grace Bio-Labs, OR, USA) onto a glass slide and a coverslip was placed on top of the spacer. The coverslip was sealed using clear nail varnish.

A mixture of 4 μm-sized and 10 μm-sized yellow-green fluorescent beads (FluoSpheres™ Sulfate Microspheres, 4.0 μm and FluoSpheres™ Polystyrene Microspheres, 10 μm, ThermoFisher Scientific, MA, USA) were used to demonstrate the approach. A mixture of 4 μm-sized and 10 μm-sized yellow-green fluorescent beads was dropped in warm 1% agarose gel solution and stirred thoroughly. Then 25 μl of the mixture was dropped in a pre-holed spacer (120 μm thick) onto a glass slide, and a coverslip was placed on top of the spacer. The coverslip was sealed using clear nail varnish. The slide was left to cool down before the experiment to solidify.

**Preparation of mouse tissues**
We used a prepared slide of sectioned mouse kidney (F24630, Invitrogen, Carlsbad, CA, USA) to demonstrate the utility of the pattern-illuminated TF. The slide contains a 16 µm cryostat section of mouse kidney stained with Alexa Fluor 488 wheat germ agglutinin, Alexa Fluor 568 phalloidin, and DAPI. While DI water was used as immersion medium for non-scattering case, and 0.15% lipid solution was used as immersion medium to mimic scattering environment since the sectioned mouse kidney is only 16 µm thick.

**Animal procedure and muscle tissue preparation**
The animal procedure (transcardial perfusion) was approved by the Massachusetts Institute of Technology Committee on Animal Care and meets the NIH guidelines for the care and use of vertebrate animals. Mice were deeply anesthetized with 1.25% avertin solution (350 mg/kg intraperitoneal) and transcardially perfused with phosphate buffered saline (PBS) containing 4% paraformaldehyde. After perfusion, thigh muscle was excised and post-fixed in 4% paraformaldehyde overnight. Muscle tissue was cryoprotected in 30% sucrose for 48 hours, embedded in Optical Cutting Temperature formulation (OCT, Tissue Tek), frozen at -20 degrees Celsius, and sliced at a thickness of 200 µm on a cryostat. Frozen sections were immersed in PBS for staining.

**Dye preparation**
Solutions of Alexa Fluor 568 Phalloidin (Invitrogen, catalog number A12380) and Hoechst 33342 (Invitrogen, catalog number H21492) were prepared as follows. The entire contents of the 300-unit vial of Alexa Fluor 568 Phalloidin was dissolved in 1.5 mL methanol to produce a 40x stock solution, which was diluted 1:40 in PBS for staining. A Hoechst stock solution was prepared by dissolving Hoechst in water at a concentration of 10 mg/mL. The Hoechst stock solution was diluted 1:2000 in PBS for a final concentration of 5 µg/mL for staining.

**Staining**
Muscle slices were permeabilized in a solution of 1% Triton-X-100 in PBS for 20 minutes at room temperature with gentle shaking. Slices were then incubated in a working solution of the dyes (dissolved in PBS as described above) for 20 minutes at room temperature with gentle shaking. Excess dye was removed by washing slices in PBS 3 times (6 minutes per wash, with gentle shaking at room temperature). Slices were then mounted on slides using Fluoromount-G or Vectashield as mounting media. Slides were coverslipped, and slides containing Vectashield as the mounting medium were sealed along the edges of the coverslip with clear nail polish. Slides were allowed to dry for at least 48 hours before imaging.

**Pattern projection**
Arbitrary pattern images were generated in Matlab by using the 'rand' function (0 or 1) for each pixel. In beads and mouse kidney experiments, patterns of 1024 × 1024 pixels resized by a factor of 8 were used for excitation patterns. This combination defines the unit block of 8 × 8 pixels at the DMD, which corresponds to 60.8 µm for the length of one side. The corresponding size of the unit block at the sample plane is 0.83 µm, which is close to the effective diffraction limit of the system $[\lambda/(2NA^2)]$. The total number of patterns for each imaging session was 256 for a complete basis set. For mouse muscle imaging, patterns of 1600 × 1600 pixels at the DMD were used to enlarge the FOV of the system with the modification of a magnifying lens compound in front of the EMCCD camera. The exposure time of the camera was adjusted in the range of 100 – 500 ms per pattern depending on the signal intensity of the specimen. The EM gain of the camera was set to be 3 – 100 depending on the signal intensity of the specimen as well.

**DEEP-TFM image reconstruction**

The calibration experiment gives the ensemble of patterns, $\{\widetilde{H}_t\}$, used to modulate spatial features. The imaging experiment gives the ensemble of measurement images, $\{Y_t\}$. Then the reconstruction of the de-scattered image, $X$, is possible from solving the set of Eq1 equations (or the set of Eq2 equations) in the main text. However, the constituting set of equations in Eq1 (and its corresponding frequency domain form in Eq2) are non-linear but quadratic with respect to the unknowns ($X$ and $sPSF$). To solve this system, one could first assume a form for $sPSF$ and then the Eq1 becomes a linear system that can be solved for $X$ with commonly available linear-optimization methods. When a solution for $X$ is found that can be substituted in Eq2 which makes a similar linear system that can be solved for $\mathcal{F}sPSF$ (and hence for $sPSF$). Thus, a proper solution for $X$ can be iteratively found. For the results shown in this paper, we only performed one iteration assuming a canonical form for $sPSF$, which resulted in visually accurate reconstructions. We used two-step iterative shrinkage/thresholding algorithm (TwIST)[15] to solve the above linear equations. Please refer to the supporting text for a detailed description of the problem formulation.